\documentclass[aps,prb,showpacs,superscriptaddress,10pt,twocolumn]{revtex4-1}

\usepackage{amsmath}

\usepackage[load-configurations=abbreviations,detect-display-math=true]{siunitx}
\usepackage{graphicx}
\usepackage{overpic}

\newcommand*\diff{\mathop{}\!\mathrm{d}}

\usepackage{xspace}

\usepackage{commath}

\def\Tc{$T_\text{c}$\xspace}
\def\Hperp{$H_{\rm c2}^\perp$\xspace}
\def\Hpar{$H_{\rm c2}^\parallel$\xspace}
\def\Hctwo{$H_{\rm c2}$\xspace}
\def\so{$\lambda_\text{so}$\xspace}
\def\ep{$\lambda_\text{sc}$\xspace}
\def\gh{$\gamma_\text{H}$\xspace}
\def\Hczero{$H_{\rm c2}(0)$\xspace}
\def\m{$\mu_0$}

\def\Hcone{$H_\text{c1}$\xspace}

\def\mTc{T_\text{c}}
\def\mHperp{H_{\rm c2}^\perp}
\def\mHpar{H_{\rm c2}^\parallel}

\def\mep{\lambda_\text{sc}}
\def\mgh{\gamma_\text{H}}
\def\mHcpar{H^\parallel_\text{c1}}
\def\mHcperp{H^\perp_\text{c1}}
\def\mHctwo{H_{\rm c2}}

\begin{document}

\title{The upper critical field of NaFe$_{1-x}$Co$_x$As superconductors}

\author{S. Ghannadzadeh}
\email{s.ghannadzadeh1@physics.ox.ac.uk}

\author{J. D. Wright}
\author{F. R. Foronda}
\affiliation{Clarendon Laboratory, Department of Physics, University of Oxford, Parks Road, Oxford, OX1 3PU, UK}

\author{S. J. Blundell}
\affiliation{Clarendon Laboratory, Department of Physics, University of Oxford, Parks Road, Oxford, OX1 3PU, UK}

\author{S. J. Clarke}
\affiliation{Department of Chemistry, University of Oxford, South Parks Road, Oxford, OX1 3QR, UK}

\author{P. A. Goddard}
\affiliation{Department of Physics, University of Warwick, Gibbet Hill Road, Coventry, CV4 7AL, UK}

\begin{abstract}

The upper critical field of NaFe$_{1-x}$Co$_x$As was measured from $x=0$ to $x=0.08$, with the magnetic field applied parallel (\Hpar) and normal (\Hperp) to the planes. The data were fitted using one-band and two-band models. The orbital and paramagnetic components of the upper critical field were extracted for \Hpar, which we find to to be strongly dominated by paramagnetic pair-breaking. In the parent compound the paramagnetic limit is equal to the value expected by BCS theory. However, substitution of Fe by Co leads to an enhancement above the BCS limit by a factor of approximately 1.6. In the over-doped region, we observe a significant convex curvature in \Hperp at low temperatures, which we attribute to the two-band nature of the superconductivity and the asymmetry between the two bands. Additionally, we discuss the behavior of critical field anisotropy, coherence length $\xi$, and the penetration depth $\lambda$.

\end{abstract}

%74.25.Op	Mixed states, critical fields, and surface sheaths 
%74.25.Dw 	Superconductivity phase diagrams 
%74.25.N- 	Response to electromagnetic fields 
%74.62.Dh	Effects of crystal defects, doping and substitution

\maketitle

%%%%%%%%%%%%%%%%%
%% Introduction                      %%
%%%%%%%%%%%%%%%%%
\section{Introduction}

The discovery of the superconductor LaO$_{1-x}$F$_x$FeAs\cite{Kamihara2006,*Kamihara2008} and the extended  Fe-pnictide superconductor family with critical temperatures\cite{Wang2008a} of up to 56 K has generated great interest and triggered huge efforts towards unmasking the underlying superconducting state. Theories for Cooper-pair symmetry range from conventional $s^{+}$-wave pairing, \cite{Kordyuk2012} to spin-fluctuation mediated $s^\pm$ pairing  with a $\pi$ phase change in the order parameter,\cite{Mazin2008} to triplet superconductivity  caused by ferromagnetic fluctuations,\cite{Brydon2011} to $d$-wave pairing.\cite{Chubukov2012} An important part of the experimental effort to verify or disprove such theories is measurement of the upper critical field \Hctwo, which is a fundamental property of type-II superconductors. Studying the temperature dependence of \Hctwo and the interplay between the various pair-breaking mechanisms may reveal the underlying pair-forming interaction. Furthermore, the critical field can be used to extract other fundamental properties such as the coherence length $\xi$ and the London penetration depth $\lambda$. The upper critical field and its anisotropy can also shed light on the possible multi-band nature of the superconductivity, and are sensitive to the dimensionality and the underlying electronic structure of the system.\cite{Kogan2012,Gurevich2003,*Gurevich2007}

In this paper, we study the upper critical field of the tenary ``111'' arsenide superconductor NaFe$_{1-x}$Co$_x$As. In stoichiometric NaFeAs, the superconducting phase coexists with an antiferromagnetic (AFM) phase which is stabilized by the delocalized Fe $d$-band electrons.\cite{Li2009a,Wright2012} This AFM phase is also preceded by a structural transition from a tetragonal to orthorhombic structure,\cite{Li2009a} similar to ``1111'' and ``122'' systems.\cite{Chu2009,*Drew2009} The evolution of the magnetic and superconducting states with addition of Co has been explored through low-field susceptibility, neutron scattering and x-ray diffraction,\cite{Parker2010} as well as muon-spin relaxation \cite{Parker2010,Wright2012} and heat capacity \cite{Wang2012a} measurements. The addition of Co causes a rapid suppression of the magnetic phase and the structural distortion, leading to the destruction of antiferromagnetism at $x \approx 0.025$.\cite{Wright2012} Substitution by Co above 3\% leads to suppression of the superconducting state itself, see Fig. \ref{fig:JW48}(a). The exact role of Co substitution in iron-pnictides is under debate, in particular it is unclear if it leads to electron doping and a rigid shift in the chemical-potential,\cite{Jiang2012,Anand2013a} or whether its primary effect is to introduce a random impurity potential leading to charge-carrier scattering,\cite{Levy2012,Allan2013,Prando2013,*Suzuki2013} or a combination of both. We note that recent studies\cite{Levy2012,Allan2013} of Ca(Fe$_{1-x}$Co$_x$)$_2$As$_2$ have shown the effect of Co substitution to be beyond a simple band-shift. However, for consistency with literature, we shall use `doping' nomenclature to refer to substitution of Fe by Co. 

Until now there has been little exploration of the critical-field behavior of this system. Recently there has been a critical-field measurement of NaFe$_{1-x}$Co$_x$As with $x=0.025$, limited to low fields and temperatures near the transition temperature \Tc.\cite{Spyrison2012} Here, we investigate the upper critical field of NaFe$_{1-x}$Co$_x$As up to 45 T, across the whole phase diagram from the under-doped to over-doped regime, including the parent compound. We find that a multi-band model is required to fully reproduce the critical-field behavior, with the fitted coupling values being insensitive to the pairing symmetry. Moreover we find the in-plane critical field to be paramagnetically limited, being enhanced over the BCS value in doped samples by a combination of spin-orbit and strong-coupling effects.

%%%%%%%%%%%%%%%%%
%% Crystal Growth                %%
%%%%%%%%%%%%%%%%%
\section{Experimental Details}
% Figure: PDO Field Sweeps
\begin{figure}[!bt]
\center
\includegraphics[width=0.9\columnwidth]{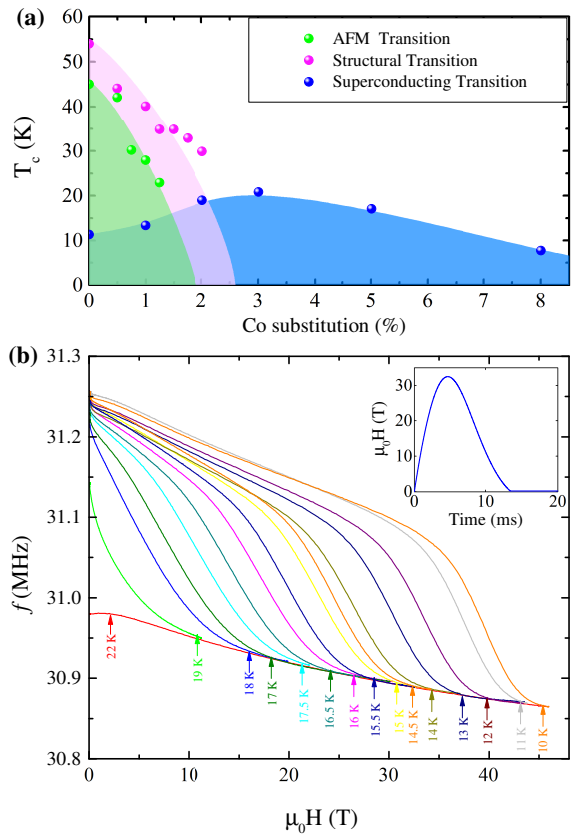}
\caption{(color online) (a) Phase diagram showing the superconducting phase as measured in this paper, together with the antiferromagnetic and the structural transitions.\cite{Wright2012} (b) Resonance frequency for $x=0.02$, at different temperatures and with the applied field parallel to the planes. The arrowed numbers indicate the temperature. The superconducting transition is seen as a drop in frequency (see main text) Inset: profile of a typical magnetic-field pulse. }\label{fig:JW48}
\end{figure}

\begin{figure*}[!tb]
\center
  \begin{overpic}[width=0.95\textwidth]{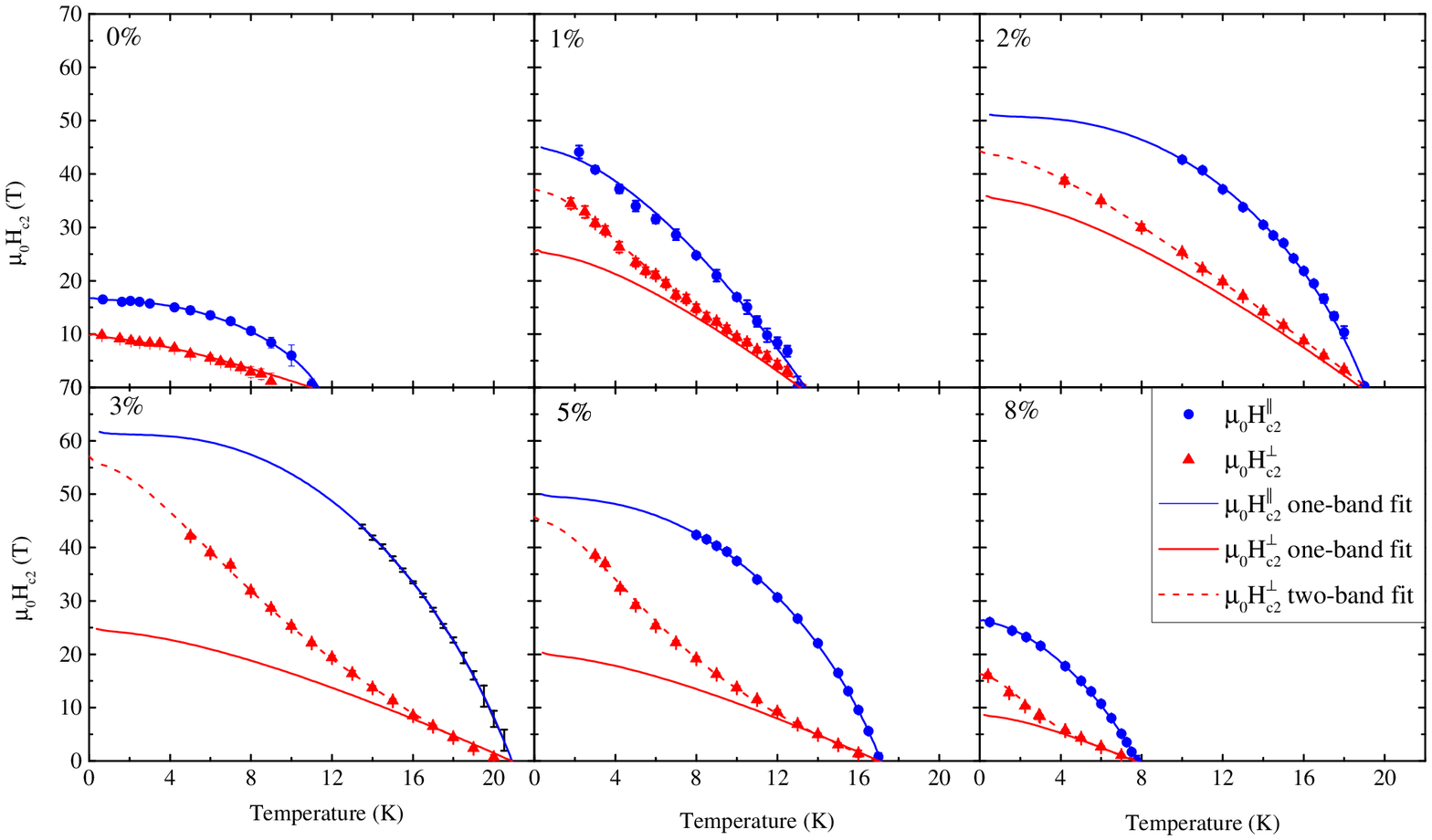}
  \end{overpic}
\caption{(color online) The upper critical field of NaFe$_{1-x}$Co$_x$As, where the percentage denotes the amount of Co substitution. Red triangles and blue circles represent $H_{\rm c2}$  with the field applied parallel and normal to the Fe-As planes, respectively. The solid lines are fits using the one-band model, and the dashed lines are fits using the two-band model (see text) with $\lambda_{11}=\lambda_{22}=1, \lambda_{11}\lambda_{22} - \lambda_{12}\lambda_{21}=0.5$.\label{fig:HC2}}
\end{figure*}

Single crystals of NaFe$_{1-x}$Co$_x$As  were synthesized using the self-flux method. Freshly cut pieces of Na were mixed with ground As in a 1:1 molar ratio and heated in a 9\,mm Nb tube (sealed under argon) to $300^{\circ}{\rm C}$ for 6 hours. The resulting black powder of overall composition NaAs was then ground with Fe$_{1-x}$Co$_x$ mixture in the ratio NaAs:Fe$_{1-x}$Co$_x$ = 2:1. This was then placed into an 8\,mm alumina crucible, itself then placed into a second sealed Nb tube and this in turn sealed in an evacuated silica ampule to prevent the oxidation of Nb. The ampule was then heated to $900^{\circ}{\rm C}$ at a rate of $\sim 1^{\circ}{\rm C}/{\rm min}$ before being slowly cooled to $400^{\circ}{\rm C}$ at a rate of $5^{\circ}{\rm C}/{\rm hr}$, at which point it was cooled rapidly to room temperature. The resulting crystals were around $1\,\times\,1\,\times\,0.1$\,mm$^3$ in dimension and could be easily removed from the flux mechanically. The growth procedure used here was found to be highly robust: the pre-reaction of Na and As minimized the reaction of Na with the alumina crucible, and sealing the crucible in Nb prevented any sublimed Na from attacking the silica ampule. X-ray diffraction measurements were carried out using an Oxford Diffraction SuperNova, which showed that no detectable impurity phases were present. Magnetic susceptibility data returned values of $T_{\rm c}$ in agreement with those previously published for each composition.\cite{Wright2012,Parker2010} The superconducting volume fraction was also in agreement with that of previous powder-sample publications.\cite{Wright2012}

%%%%%%%%%%%%%%%%%
%% PDO                                    %%
%%%%%%%%%%%%%%%%%

$H_{\rm c2}$ was measured in pulsed magnetic fields at the Nicholas Kurti Magnetic Field Laboratory, UK. The proximity detector oscillator technique was used, in which the sample is placed inside a resonator coil and the resonance frequency is monitored as function of field and temperature.\cite{Altarawneh2009,Ghannadzadeh2011} This frequency depends strongly on the sample skin/penetration depth, and so, due to the difference between the superconducting penetration depth and normal state skin depth, the superconducting-normal state phase transition can be seen as a large change in the resonant frequency at the critical temperature or field. One can find the upper critical field $H_{\rm c2}$, as opposed to the irreversibility field, by setting $H_{\rm c2}$ to be the field at which the frequency (at $T <T_c$) matches that of the  normal state at $T > T_c$, for example see Fig. \ref{fig:JW48}(b).\footnote{\label{footnote:Hc2}One usually draws a linear extrapolation of the transition curvature, and sees where it crosses the normal state line to find \Hctwo.\cite{Altarawneh2008} In order to be more systematic, we define \Hctwo as the peak in $\diff^2 f/\diff H^2$. Both methods give similar results.} Phase transitions determined by this technique have been shown to be consistent with those measured by  traditional resistive contact techniques.\cite{Mun2011} The critical field was measured with the magnetic field applied  parallel ($H_{\rm c2}^\parallel$) or normal ($H_{\rm c2}^{\perp}$) to the Fe-As planes.  The samples were covered with dried Apiezon N grease to prevent sample degradation. It was found that sample degradation leads to a noticeable change in \Tc and a broadening of the transition, with the parent compound being the most sensitive. Therefore \Tc and the shape of transition were monitored carefully to detect sample degredation. We observed no heating in the pulsed magnetic-field measurements.\footnote{Heating can arise due to the high $\text{d}B/\text{d}T$. We checked for heating by monitoring \Hctwo as the pulsed width was changed. We observed no change in \Hctwo for pulse widths above 10~ms.}

%%%%%%%%%%%%%
%% Results %%
%%%%%%%%%%%%%
\section{Results}
The upper critical fields for NaFe$_{1-x}$Co$_x$As with 0\%, 1\%, 2\%, 3\%, 5\% and 8\% Co doping are given in Fig. \ref{fig:HC2}. In the parent compound, we observe a critical temperature of 11~K  and \m\Hperp (\m\Hpar) of 9.9~T (16.6~T) as the temperature tends towards absolute zero. Doping slightly to 1\% leads to a significant increase in \Hctwo, with the critical fields in both directions increasing linearly as the temperature is decreased. This is in contrast to the concave temperature dependence of the parent compound. Doping to 2\% causes \Hpar to become noticeably concave, increasing the difference between the upper critical field of the two orientations. At 3\% we find the highest $T_\text{c}$ of 21~K, as well as the highest critical fields. Interestingly, at optimal doping \Hperp begins to rise upwards as the temperature is increased and becomes convex, in contrast to the behavior of \Hpar.  Further addition of Co moves us to the over-doped regime, where any additional doping leads to a decrease in $T_\text{c}$ and the upper critical fields, although \Hpar and \Hperp retain their shape. Thus we notice two overall trends. One is the usual increase, and then decrease, of \Hctwo  as we move over from the under-doped to over-doped regime, see Fig.  \ref{fig:doping}(a). The other is the gradual change in the behavior of \Hperp from one extreme to another: from very concave at parent composition, to linear at 1\% doping, to convex at and above optimal-doping. We also note that our data agree well with the previous low-field resistivity study of  2.5\% doped samples.\cite{Spyrison2012}

%%%%%%%%%%%%%
%% Limiting Field %%%
%%%%%%%%%%%%%
\section{Discussion}

The upper critical field is determined by two independent mechanisms, one of which is the Clogston-Chandrasekhar spin paramagnetism.\cite{Chandrasekhar1962,*Clogston1962}  The spin paramagnetism reduces the free energy  of the normal state, thus in effect reducing the superconducting energy gap. The paramagnetic (PM) ƒlimit is defined as $\mu_0 H^\text{pm}_\text{c2}(0)= {\Delta}/(\sqrt{g} \mu_B)$, where $\Delta$ is the zero-temperature energy gap and $g$ is the $g$-factor. In the case of isotropic BCS $s$-wave pairing with $g=2$, this reduces to $H^\text{pm}_\text{BCS}(0) \approx 1.84 T_c$. However, this weak BCS coupling can underestimate the actual paramagnetic limit. The critical field can be enhanced by an increase in $\Delta$, for example by strong electronic interactions giving $H^{\rm pm}_\text{c2} = (1 + \mep)H^\text{pm}_\text{BCS}$ ($\mep$ is the strong-coupling coefficient), or by spin-orbit coupling, resulting in $g < 2$. Additionally, any spin-orbit or spin-flip scattering destroys spin as a good quantum number \cite{Werthamer1966} and leads to depairing of the Cooper pair, thus bringing  the spin paramagnetism of the superconducting and normal states closer together and reducing the limiting effect of paramagnetism on the critical field.\cite{Werthamer1966,Cody1968,Maki1966} The second mechanism which limits the critical field is orbital pair breaking, which is related to the creation of Abrikosov vortex lines and superconducting currents around the vortex core, reducing the superconducting condensation energy. In general $H_\mathrm{c2}$ is influenced by both orbital and spin-paramagnetic effects (although usually one tends to dominate), and the relative importance of these effects can be measured by the Maki parameter $\alpha = \sqrt{2} H^\text{orb}_\text{c2} /H^\text{pm}_\text{c2}$.\cite{Maki1966,Werthamer1966}

Angle resolved photoemission spectroscopy measurements\cite{Liu2012,Cui2012,Liu2011} and band structure calculations\cite{Deng2009,Kusakabe2009} show NaFe$_{1-x}$Co$_x$As to be a quasi-two-dimensional superconductor with a Fermi surface consisting of collinear warped columns. In such systems, application of magnetic field parallel to the superconducting planes causes the electrons to form open orbits along the cylinders,\cite{Singleton2007} which leads to a negligible orbital effect and significant enhancement of the orbital critical field above what would be expected for a more isotropic superconductor. In contrast, application of field normal to the superconducting planes forms closed electron orbits within the Fermi sheets and leads to creation of vortices,\cite{Singleton2002,*Nam1999a} thus reducing the orbital limit.\cite{Lee2009} Therefore in general $\mHpar > \mHperp$, as is the case for NaFe$_{1-x}$Co$_x$As (see Fig. \ref{fig:HC2}).

In this paper we assume that the samples are in the dirty limit, where mean free path $l$ $\ll$ coherence length $\xi$. This is justified by the low residual resistivity ratio (as low as $4$),\cite{Spyrison2012} the low superconducting volume fraction ($< 10\%$) of the parent compound,\cite{Wright2012} and the tendency of pnictide superconductors to have a small $l$ due to the low Fermi velocity.\cite{Weickert2011} Furthermore, we will assume that the system is weakly coupled; however we note that NaFe$_{1-x}$Co$_x$As heat capacity data may be understood using Eliashberg theory in the strong-coupling limit, albeit with significant modifications.\cite{Tan2013}

\subsection{Single-band model}

%%%%%%%%%%%%%%%%
%% WHH Model             %%%
%%%%%%%%%%%%%%%%

One model for the upper critical field of a superconductor is the single-band Werthamer-Helfand-Hohenberg (WHH) model in the dirty limit.\cite{Werthamer1966} This model includes the effects of paramagnetism, orbital pair-breaking, and spin-orbit scattering; however it neglects any strong coupling. $H_{c2}$ is given implicitly via
\begin{align*}
\ln{\frac{1}{t}} = &\sum^\infty_{-\infty} \left\{ \frac{1}{|2v+1|} \right. \\
 & \left. - \left[ |2v+1| + \frac{h}{t} + \frac{(\alpha h t)^2}{|2v+1|+(h+ \lambda_\text{so})/t} \right]^{-1} \right\},
\end{align*}
where $t= T/T_c$, $h= 4 H_{c2} / \left[ \pi^2 T_c (\diff H_{c2}/\diff T)_{T=T_c} \right]$, $v$ is the summing index and $\lambda_\text{so}=\hbar/(3 \pi k_\text{B} T_\text{c} \tau_\text{so})$ accounts for the spin-orbit and spin-flip scattering with $\tau_\text{so}$ as the mean free scattering time.\cite{Cody1968} The results of fits to the WHH model are shown in Fig. \ref{fig:HC2} (solid lines). \Hpar is well described by this model across all temperatures, however in contrast \Hperp seems to be only well-fitted close to \Tc --- the WHH model is insufficient to reproduce the enhancement  of the upper critical fields at low temperatures. 

%%%%%%%%%%%%%%%%%%%%%%%
%% Upwards curvature in H, etc.      %%%
%%%%%%%%%%%%%%%%%%%%%%%

One possibility for such a convex curvature in $H_\text{c2}(T)$ is localization of the electron wave functions,\cite{Lebed1986,Dupuis1993,Kano2009}  where the application of high magnetic fields below a critical temperature $T^*< T_\text{c}$ leads to the complete confinement of  electronic motion to the plane of the applied magnetic field, such that the net magnetic flux experienced by the confined Cooper pairs is zero. This suppresses orbital pair breaking, and in the absence of other pair-breaking effects leads to an increase in the critical field and the possibility of re-entrant superconductivity at high magnetic fields. However, this would imply that the convex curvature would be seen in \Hpar, whereas we have an upwards curvature in \Hperp instead.  The second possibility, in the case of weak interlayer coupling,  is the formation of a Josephson-coupled superconductor with flux lines trapped within layers.\cite{Klemm1975} This causes a divergence in the orbital critical field and leads to a rapid increase of \Hctwo if the system is orbitally limited. However, this phenomenon can also be ruled out since (as shown later) the superconducting coherence length is found to be larger than the interlayer separation.

\begin{figure}
	\centering
	\includegraphics[width=.9\columnwidth]{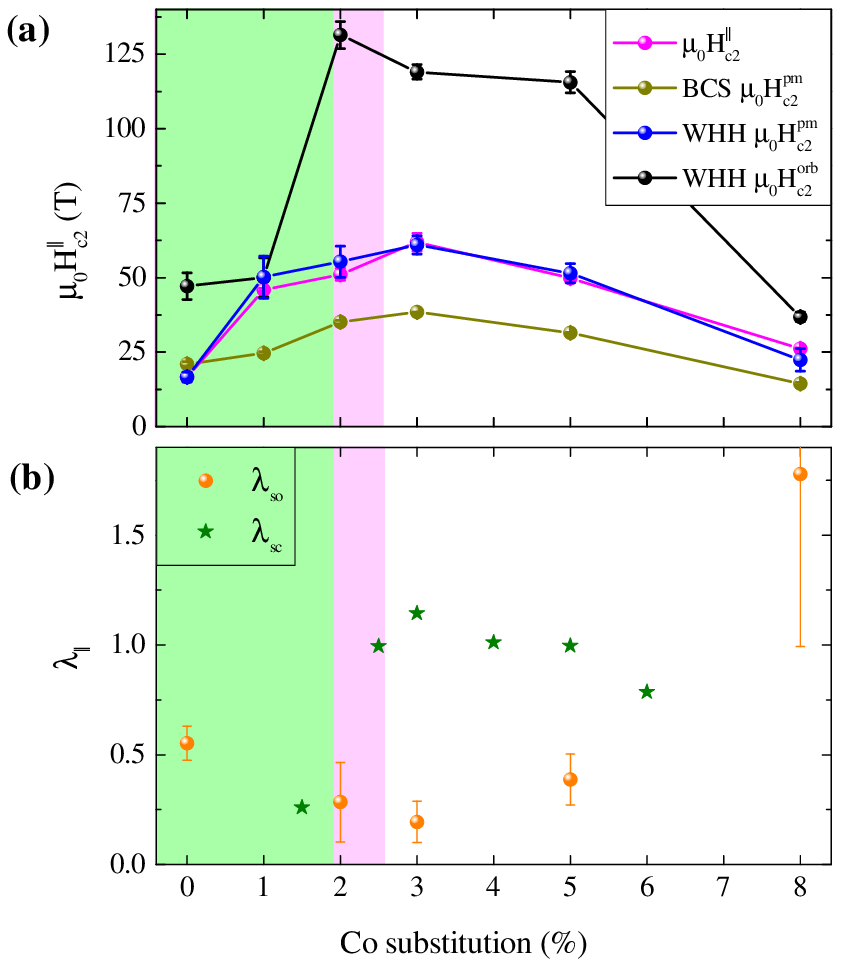}
	\caption{(color online) Evolution with Co doping of (a) \Hpar and the BCS paramagnetic limit, as well as the orbital and paramagnetic components deduced from fits to the WHH model, (b) the spin-orbit scattering constant \so for \Hpar (for 1\%, a reliable estimation is not possible) and the strong-coupling constant \ep (extracted from Ref. \onlinecite{Tan2013}).\label{fig:doping} The green, purple and white areas represent the orthorhombic AFM phase, the non-magnetic orthorhombic phase, and the tetragonal phase, respectively.}
\end{figure}

\subsection{Two-band model}
It is likely that the upwards curvature in \Hperp is due to the two-band nature of the superconductivity, as is thought to be the case for the rare-earth 1111 systems ({\it Re}FeAsO),\cite{Hunte2008,Jaroszynski2008,Lee2009} the 122 systems (BaFe$_2$A$_2$),\cite{Kano2009} and the closely related 111 superconductor LiFeAs,\cite{Hunte2008} amongst others. A two-band superconductor can be fitted using the model by Gurevich,\cite{Gurevich2003} which is based on a weakly-coupled system in the dirty-limit\footnote{The use of the dirty-limit does not matter in this case, since explaining the large negative curvature in the clean-limit also requires\cite{Dahm2003} a significant band asymmetry.} and whose parameters are the BCS coupling constants $\lambda_{nm}$ ($\lambda_{12}$, $\lambda_{21}$ interband; $\lambda_{11}, \lambda_{22}$ intraband) and the band diffusivities $D_1$ and $D_2$. This model accounts well for our \Hperp as shown in Fig. \ref{fig:HC2}, and attributes the enhancement of \Hperp as well as the convex curvature to a significant difference between the diffusivity of the two bands, $D^\perp_2/D^\perp_1 < 1$.\cite{Gurevich2003,*Gurevich2007} The $D$ anisotropy could be due to one band being dirtier than the other,\cite{Gurevich2007} scattering from magnetic impurities or strong magnetic excitations,\cite{Jaroszynski2008,Lee2009} or a difference between the hole and electron Fermi velocities ($\upsilon_\text{e} \gg \upsilon_\text{h}$).\cite{Deng2009} We note that in contrast to \Hperp, the \Hpar data can be fitted by setting $D^\parallel_1=D^\parallel_2$,\footnote{$D^\parallel$, the effective diffusivity when the field is applied parallel to planes, is defined as $D^\parallel = \sqrt{({D^\perp}^2 + D^\perp D^{(c)})}$ } which reduces back to a one-band model and thus validates our previous \Hpar one-band fit.\cite{Gurevich2003,*Gurevich2007} This model was found to be strongly sensitive to the band diffusivities $D_1$ and $D_2$, but significantly less dependent on the BCS couplings.  Furthermore, due to the symmetry of the Gurevich functions our data can be fitted with attractive interband coupling $\lambda_{12},\lambda_{21}>0$ indicative of the conventional $s$-wave coupling scenario, as well as with repulsive interband coupling $\lambda_{12},\lambda_{21}<0$ as would be the case for $s^\pm$  superconductivity.\cite{Mazin2008} Thus our fit could be compatible with the recent neutron study showing support for $s^\pm$ coupling.\cite{Zhang2013}  Both scenarios could be fitted  with strong ($\lambda_{12}\lambda_{21} > \lambda_{11}\lambda_{22}$) or weak ($\lambda_{12}\lambda_{21} < \lambda_{11}\lambda_{22}$) interband coupling. Therefore we find the two-band Gurevich model to be over-parameterized in our situation, and so we shall not rely on this model for extracting values for $\lambda_{nm}$. 

There also exists the possibility of a Fulde-Ferrell-Larkin-Ovchinnikov (FFLO) phase,\cite{Fulde1964,*Larkin1964} which would lead to a state of inhomogeneous superconductivity with a spatially varying gap function, causing an increase in the experimentally measured critical field below $0.55\mTc$ if the system is paramagnetically limited and hence accounting for the slope of \Hpar.\cite{Matsuda2007} The presence of an exotic FFLO phase cannot be ruled out, however it seems unlikely since (a) such an state is possible only in extremely clean materials ($l\gg \xi$), and (b) that the FFLO phase is thought to cause a hysteresis between the up-sweep and down-sweep measurements.\cite{Lortz2007,Adams1998} However, we note that the possibility of a weak FFLO state has been raised for LiFeAs,\cite{Cho2011} where $l \approx \xi$.\cite{Okada2012} 

%%%%%%%%%%%%%%%%
%% PM Enhantment   %%%
%%%%%%%%%%%%%%%%
\subsection{Enhancement of the paramagnetic limit}
Given that in our case the Gurevich model collapses into a one-band model when the applied field is parallel to the planes, we return to further analysis of the single-band fit of the WHH model to \Hpar, which allows us to separate the orbital and paramagnetic components of the critical field. The orbital limit at \Tc can be approximated as $H^\text{orb}_\text{c2}(0) = - 0.69 T_c (\diff H^\parallel_{c2}/\diff T)_{T=T_c}$ in the dirty limit, and the paramagnetic component can be found via $H^\text{pm}_\text{c2}(0) = \sqrt{2} H^\text{orb}_\text{c2}(0)/\alpha$. The extracted orbital and paramagnetic components, the BCS paramagnetic limit $\mu_0 H^\text{pm}_\text{BCS}(0) \approx 1.84 T_c$, and the experimental critical field \Hczero (as extrapolated using the WHH model), is given in  Fig. \ref{fig:doping}(b). We observe that at all dopings the extracted paramagnetic limit matches $H^\parallel_{c2}(0)$ well, while the calculated orbital limit is noticeably higher than $H^\parallel_{c2}(0)$. This implies that the critical field behavior is dominated by paramagnetic effects. The only exception is at 1\% doping, where $H^\text{orb}_\text{c2}(0)\approx H^\text{pm}_\text{c2}(0)$ such that the effective critical field may be a function of both effects. For the parent compound we observe the PM limit to be equal to that expected by BCS theory, however addition of Co leads to a significant enhancement of the PM limit above $H^\text{PM}_\text{BCS}$ by about 1.6. This enhancement of the PM limit is seen in all doped samples. Paramagnetic enhancement above the BCS limit seems to be a  feature of Co-doped pnictides,\cite{Khim2011} and it would be interesting to compare this to the more isotropic (Ba,Sr)Fe$_{2-x}$Co$_{x}$As$_2$,\cite{Yuan2009,Khim2010} which show paramagnetic enhancement by about $\sim $ 4 -- 5 and are thus likely to be orbitally limited instead.

The spin-orbit scattering constant  \so, as extracted from the fit of the WHH model to \Hpar, is shown in Fig. \ref{fig:doping}(b). Starting from the parent compound, \so decreases upon doping, due a the reduction in scattering from magnetic excitations as the system moves away from the long-range ordered AFM phase. \so then reaches a minima at optimal doping, but begins to increase in the over-doped region, possibly due to scattering from magnetic Co impurities. It is likely that the enhancement of the paramagnetic limit is due to a combination of the spin-orbit scattering and other strong electronic interactions; any increase in the $g$-factor or the effective mass caused by such interactions will enhance the paramagnetic limit by a further $(m^* g)/(mg^*)$.\cite{Perez-Gonzalez1996,Schossmann1989a, Zuo2000} We note that neither of the models considered so far include the Eliashberg strong-coupling renormalization effects.\cite{Carbotte1990} Due to its two-band nature, NaFe$_{1-x}$Co$_x$As has two band gaps of different magnitude. Heat capacity measurements by \citet{Tan2013} have shown the larger band gap $\Delta_\text{L}(0)$ to be significantly above the BCS value $\Delta_\text{BCS}(0) = 3.5 k_\text{B} \mTc /2$, and allow us to extract the effective strong-coupling constant \ep via $\Delta_\text{L}(0) = (1+\mep)\Delta_\text{BCS}(0)$.  This is shown in Fig. \ref{fig:doping}(b). We have a maximum $\mep = 1.14$ at optimal doping, implying the presence of significant  strong  electronic interactions, which are reduced as we move away from optimal doping. Therefore we conclude that the enhancement of the paramagnetic limit above the BCS value is a combination of both the spin-orbit and strong coupling effects. However, the reason for the absence of this PM enhancement in the parent compound, despite a finite \so, is not clear. Since $\mep \approx 0$ in the parent compound, this may indicate that strong coupling effects dominate the PM enhancement effect; but in such a case one wonders why there is a PM enhancement at 1\% doping (which also has $\mep \approx 0$). Alternatively, the absence of PM enhancement in the parent compound may be caused by some underlying interaction between the magnetism and superconductivity, caused by the strengthening of the AFM phase in the 0\% doping region. It may also indicate that the PM limit is influenced by additional effects that are not encapsulated in \so or \ep.

\begin{figure}
	\centering
 \includegraphics[width=.85\columnwidth]{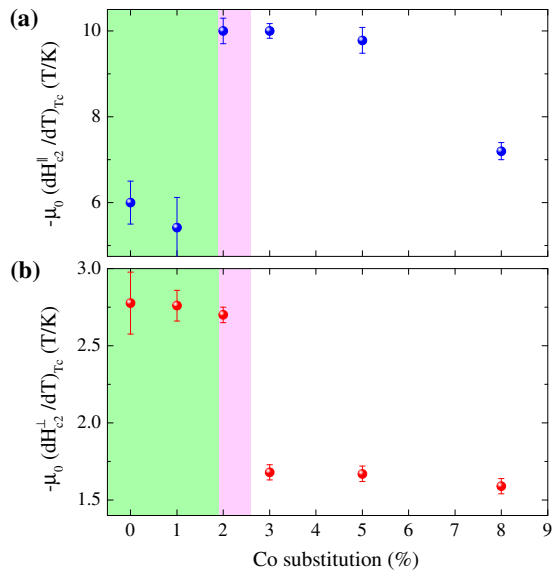}
	\caption{(color online) Differential with respect to temperature of (a) $\mu_0 \mHpar$ and (b) $\mu_0 \mHperp$, at \Tc and as a function of doping. The green, purple and white areas represent the orthorhombic AFM phase, the non-magnetic orthorhombic phase, and the tetragonal phase, respectively.  \label{fig:dHdT}}
\end{figure}

As explained previously, application of field parallel to the crystal planes leads to a significant $H^\text{orb}_\text{c2}$, thus allowing the paramagnetic effect to limit the critical field. It is this reduction in the importance of the orbital character which allows the one-band WHH model to describe the \Hpar despite the underlying two-band nature of the system, and which sets the scene for the dominance of the paramagnetic limit. Within this regime we find the WHH model to be a better fit, as it includes the spin-orbit effects described previously. In contrast, application of field normal to the planes reduces the orbital critical field, and so the overall critical field now becomes an intricate combination of the two-band orbital and paramagnetic limits. This results in the one-band WHH model being no longer sufficient due to the increased sensitivity of the critical field to the underlying multi-band nature of the Fermi surface.

%%%%%%%%%%%%%%%%
%% dHc/dT                        %%%
%%%%%%%%%%%%%%%%

\subsection{$(\text{d}\mHctwo /\text{d}T)_{T=\mTc}$}

Even though \Hperp cannot be fully described by a one-band model, $-\mu_0 (\diff \mHperp /\diff T)_{T=\mTc}$ at \Tc is still of interest as it should be proportional to the orbital critical field of the cleaner band.\cite{Gurevich2007} The $-\mu_0 (\diff H_\text{c2} /\diff T)_{T=\mTc}$ for each sample orientation is plotted in Fig. \ref{fig:dHdT}, where we notice a sudden change at 2\%-3\% doping for both. It is at this doping regime that the the structural transition and the antiferromagnetism become suppressed, leading to the possible formation of a quantum critical point at absolute zero. The AFM structure is thought to form a $(\pi,0)$ columnar spin-stripe structure within the Fe-As planes, with simple columns of up/down spins along the $c$-direction.\cite{Li2009a,Fernandes2010} The coherence length is several times the relevant Fe-Fe separation distance $d$ in both in-plane ($\xi_\parallel \sim 7 d_\parallel)$ and out-of-plane ($\xi_\perp \sim 7 d_\perp$) directions, such that the Cooper pairs should feel a net internal field of zero. This implies that the sudden change in $(\diff H_\text{c2} /\diff T)_{T=\mTc}$, and hence the corresponding effect on the orbital limit, is not simply due to internal mean free field effects. Thus it must be due to either a more inherent interaction between the superconductivity and antiferromagnetism, or due to the structural transition. We note that the sudden change in $(\diff \mHpar /\diff T)_{T=\mTc}$ at 2\% doping coincides with the suppression of antiferromagnetism, while the drop in $-(\diff \mHpar /\diff T)_{T=\mTc}$ coincides with the structural transition at $\approx$ 3\% doping. Furthermore, we note that the sharp increase in \ep at 2\% doping also coincides with the death of antiferromagnetism. 

%%%%%%%%%%%%%%%%%%
%% Anisotorpy          %%%
%%%%%%%%%%%%%%%%%%

\subsection{Anisotropy}
\begin{figure}
	\centering
	\includegraphics[width=0.9\columnwidth]{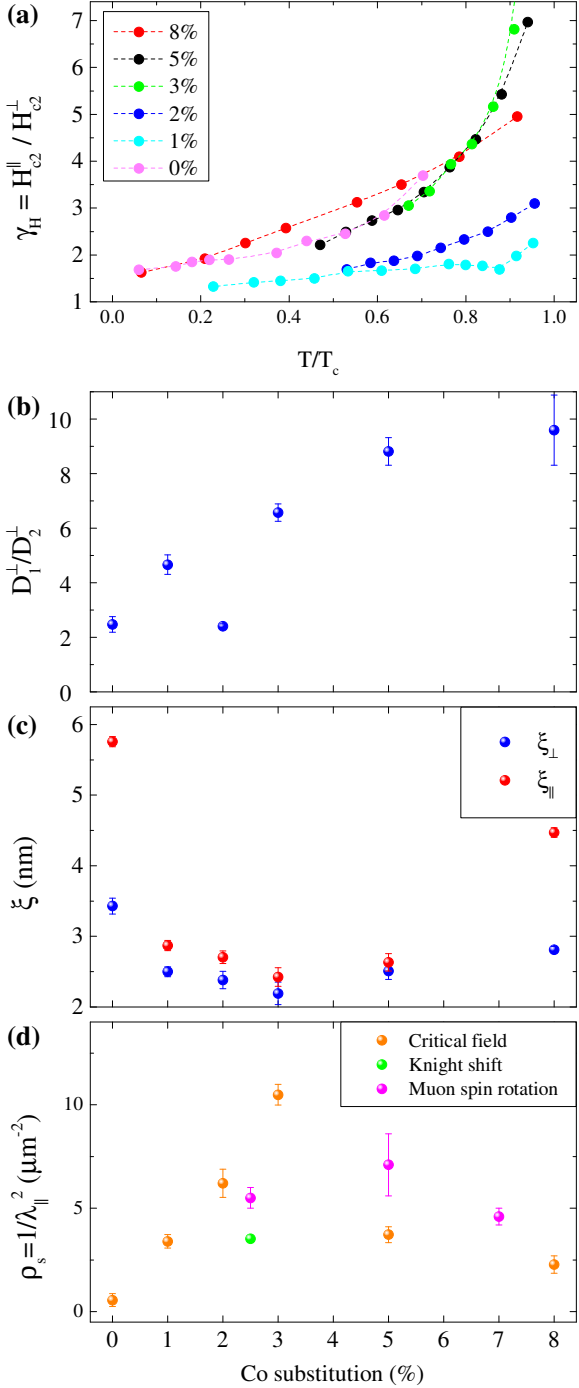}		 
	\caption{(color online) (a) The critical field anisotropy $\gamma_\text{H}$ as a function of $T/\mTc$; (b) plot of the out-of-plane band diffusivity anisotropy $D^\perp_1 /D^\perp_2$, from fits to the \Hperp with $\lambda_{11}=\lambda_{22}=1, \lambda_{11}\lambda_{22} - \lambda_{12}\lambda_{21}=0.5$. Also shown are plots of (c) coherence lengths at 0~K, and (d) evolution of the superfluid stiffness with doping at 2~K, where we have also included data from knight shift\cite{Oh2013} and muon-spin rotation\cite{Wright2012} measurements.  \label{fig:lengths}}
\end{figure}

Given the critical field in both directions, we obtain the critical field anisotropy via $\gamma_\text{H} =  \mHpar / \mHperp$, as given in Fig \ref{fig:lengths}(a). The anisotropy is highest at \Tc, and decreases with decreasing temperature. Such a temperature-dependent anisotropy is characteristic of pnictide superconductors, having being seen in Sr$_{1-x}$Eu$_x$(Fe$_{0.89}$Co$_{0.11}$)$_2$As$_2$,\cite{Hu2012} BaFe$_2$As$_2$,\cite{Kano2009} SmFeAsO$_{0.85}$,\cite{Lee2009} and LiFeAs.\cite{Zhang2011a} This is a signature of the underlying multi-band nature of the system, since in a single-band $s$-wave superconductor the anisotropy should be largely temperature independent, although we note that single-band $d$-wave superconductors may have a weakly temperature dependent \gh.\cite{Kogan2012} However, we do not see the \gh peak at $T=0.9\mTc$ as expected in a two-band system\cite{Golubov2003} and seen in a number of other materials,\cite{Lee2009,Mun2012,Kano2009} possibly due to the poor resolution near \Tc. The approach to unity at low temperatures can be explained using several effects: (a) by simply setting the paramagnetic limit  as the upper critical field at absolute zero for both directions, (b) by assuming that at low temperatures the  band with lower diffusivity anisotropy  ($D_\parallel / D_\perp$) dominates, or (c) that the two bands have opposite anisotropy which compensate each other at low temperatures, for example $D^1_\parallel / D^1_\perp<1$, $D^2_\parallel / D^2_\perp>1$.\cite{Hu2012}

We note that with the exception of the parent compound, the under-doped systems are relatively isotropic ($\mgh < 3$), whereas the optimally and over-doped systems become noticeably anisotropic near \Tc ($\mgh \approx 7$). This is complemented by the fact that the positive curvature in \Hperp also becomes more pronounced as we move over from the under-doped to the over-doped regime (Fig. \ref{fig:HC2}). Both observations could be accounted for by an increase in the band diffusivity anisotropy $D_1 /D_2$ [see Fig. \ref{fig:lengths}(b)], which magnifies the effect of the underlying multi-band nature on  \gh and \Hperp. It is known that  the NaFe$_{1-x}$Co$_x$As Fermi surface consists of hole pockets at $\Gamma$ and two electron pockets on the $M$ points of the reciprocal space.\cite{Cui2012,Liu2011} The Fermi surface is found to be highly sensitive to Co doping; causing the electron pockets to widen and hole pockets to shrink with doping, until the hole pockets completely vanish at $x \approx 0.1$.\cite{Cui2012} It is this difference between the response of the hole and electron pockets (and the corresponding change in the Fermi velocity $\upsilon_\text{F}$), as well as the increase in scattering from magnetic impurities with additional doping, which leads to an increase in $D_1 /D_2$.

%%%%%%%%%%%%%%%%%%
%% Coherence length          %%%
%%%%%%%%%%%%%%%%%%
\subsection{Coherence length and the superfluid stiffness}

The Ginzburg-Landau coherence lengths can be estimated  using $\xi_\parallel = \sqrt{\phi_0/(2 \pi \mu_0 \mHperp)}$ and $\xi_\perp = \phi_0/(2 \pi \mu_0\mHpar \xi_\parallel)$, where $\phi_0$ is the flux quantum.  Extrapolating the critical field measurements to absolute zero using the WHH model for \Hpar and the Gurevich model for \Hperp, we can calculate the coherence lengths as given in Fig. \ref{fig:lengths}(c), showing a minima at optimal doping as expected.\footnote{The uncertainly in extrapolation of the over-parametrized two-band model to 0~K has been included in error bars of Fig. \ref{fig:lengths}(c).} The superconducting layer thickness $\delta$ can be estimated as $\delta=\sqrt{12}\xi_\parallel \mHperp / \mHpar$,\cite{Schneider1993} which gives us a minimum superconducting layer thickness of \SI{7.25}{nm}, an order of magnitude longer than the lattice $c$-axis parameter.  This rules out the possibility of a dimensional crossover, or the formation of a Josephson-coupled superconductor, as discussed previously. However, it has been noted that the actual coherence lengths may be smaller than given by the Ginzburg-Landau model due to competition with paramagnetism.\cite{Kano2009}

%%%%%%%%%%%%%%%%
%% Penetration depth  %%%
%%%%%%%%%%%%%%%%

The lower critical field $H_\text{c1}$, which denotes the onset of vortex formation, was also measured using a Quantum Design SQUID magnetometer,\cite{QuantumDesign} where the sample magnetization was recorded after zero-field cooling from 300~K down to  2~K. \Hcone was determined as the point at which the magnetization deviates from the Meissner condition $\chi=-1$. Given the lower critical field, we can calculate the London penetration depth via $\mu_0 \mHcperp = (\phi_0 / 4 \pi \lambda^2_\parallel) \ln(\lambda_\parallel / \xi_\parallel)$ and $\mu_0 \mHcpar = (\phi_0 / 4 \pi \lambda_\parallel  \lambda_\perp) \ln(\sqrt{\lambda_\parallel  \lambda_\perp/ \xi_\parallel \xi_\perp})$.  The penetration depth is related to the number density $n_i$ of charge carriers and the superfluid density via $\lambda^{-2} \equiv \rho_s \propto  \textstyle{\sum} n_i$. The evolution of the superfluid density $\rho_s$ at 2~K  is given in Fig. \ref{fig:lengths}(d), where we observe a maximum at optimal doping, with $\rho_s$ decreasing almost linearly in the under-doped region. Such a rapid fall is believe to be due to the competition between superconductivity and antiferromagnetism which removes a portion of charge carriers that might otherwise  join the condensate,\cite{Gordon2010,Prozorov2011} combined with an increase in magnetic pair-breaking scattering [\so, see Fig. \ref{fig:doping}(b)] as we move deep into the AFM region.\cite{Luan2011} The slower decrease within the over-doped regime may be a result of the reduction in the carrier density as the hole pockets begin to shrink, again combined with pair-breaking scattering. This reduction in $\rho_s$ in the over-doped region can also be caused by the anisotropic superconducting gap in the $s^\pm$ coupling scenario, with an anisotropic reduction in the pairing strength as we move away from optimal doping.\cite{Luan2011,Vorontsov2009} Finally, we emphasize that the $1/\lambda^2_\parallel$ values given here are approximate as we have neglected any geometric factors as well as demagnetization effects at sharp corners of the sample,\cite{Brandt1999} however we note that our data is in relatively good agreement with previous knight shift\cite{Oh2013} and muon-spin rotation\cite{Wright2012} measurements.

%%%%%%%%%%%%%%%%
%% Conclusion                 %%%
%%%%%%%%%%%%%%%%

\section{Conclusion}

We have measured the upper critical field of NaFe$_{1-x}$Co$_x$As across the Co doped phase diagram, with the magnetic field applied normal and parallel to the superconducting planes. We find that the in-plane critical field is paramagnetically limited across the phase diagram, with the critical field matching the BCS limit in the parent compound.  Doping leads to an enhancement of the critical field above the BCS paramagnetic limit due to a combination of spin-orbit scattering and strong electronic interactions. We note that a multi-band model is necessary to fully describe the critical field behavior, in particular the upwards curvature in \Hperp at low temperature, and is further confirmed by the critical field anisotropy. We also discover a sudden change in  $(\diff H_{c2}/\diff T)_{T=T_c}$ at optimal doping, which we relate to the death of the antiferromagnetic magnetism and/or the structural transition near optimal doping. We also discuss the penetration depth $\xi$ and discover a peak in the superfluid density at optimal doping.

\begin{acknowledgments}
This work is supported by EPSRC (UK). PAG thanks the University of Oxford for the provision of a Visiting Lectureship.
\end{acknowledgments}

\bibliographystyle{apsrev4-1}
\bibliography{/Volumes/DataHD/SamanProfile/DPhil/JabRef/database}

\end{document}